# High-Rate Entanglement Source via Two-Photon Emission from Semiconductor Quantum Wells

Alex Hayat, Pavel Ginzburg, and Meir Orenstein[a]

Department of Electrical Engineering, Technion, Haifa 32000, Israel

We propose a compact high-intensity room-temperature source of entangled photons based on the efficient second-order process of two-photon spontaneous emission from electrically-pumped semiconductor quantum wells in a photonic microcavity. Two-photon emission rate in room-temperature semiconductor devices is determined solely by the carrier density, regardless of the residual one-photon emission. The microcavity selects two-photon emission for a specific signal and idler wavelengths and at a preferred direction without modifying the overall rate. Pair-generation rate in GaAs/AlGaAs quantum well structure is estimated using a 14-band model to be 3 orders of magnitude higher than for traditional broadband parametric down-conversion sources.

03.67.Mn, 42.50 Dv, 32.80.Wr

---

[a] meiro@ee.technion.ac.il

Entangled-photon states are essential in various applications of optical quantum information processing, including quantum computation, quantum cryptography and teleportation [1]. Furthermore, entanglement highlights most vividly the non-locality of quantum mechanics through violation of Bell's inequalities [2] in contrast to the local realism of classical physics [3].

The earliest attempts to produce polarization-entangled photons by 2-$\gamma$ photon decay of positronium [4] demanded strong supplementary assumptions for tests of Bell's inequalities due to the lack of high-energy photon polarizers. Pairs of low-energy photons emitted in certain atomic radiative cascades [5] yield better results; however these sources suffer from low brightness and polarization degradation caused by the atomic recoil. Solid-state with higher material density enables a significant increase in the emission rate of the source. The most popular sources of entangled photons today are based on parametric downconversion (PDC) of pump photons into signal-idler pairs in non-centrosymetric crystals with second-order optical nonlinearity [6]. The efficiency of PDC-based entanglement sources is limited, however, because of the required post-selection or spatial filtering and the relatively weak fundamental interaction. PDC is described by a third-order non-resonant process in the time-dependant perturbation theory [7] combined with a first-order process of the pump laser emission. Moreover, PDC $\chi^{(2)}$ nonlinear interaction requires dispersion compensation techniques using birefringence or quasi-phasematching [8]. Recently experiments were performed demonstrating the generation of entangled photons employing standard telecom fibers via Kerr nonlinearity [9] avoiding the output photon fiber-coupling problem and with much less severe phasematching requirements, however the underlying weaker $\chi^{(3)}$ nonlinearity still limits the source performance. Alternative approach using semiconductor quantum dots (QDs) offers sources of entangled photons on-demand [10]; nevertheless employing QD sources in quantum communications appears to be difficult due to their low generation rates and cryogenic operation.

Here we propose a simple high-efficiency room-temperature entangled photons source based on spontaneous two-photon emission (TPE) from quantum wells (QWs) in a semiconductor photonic microcavity. The calculated high rate emission is mainly due to the fact that resonant second-order TPE interaction is much more efficient than a fourth-order optically-pumped PDC process; by at least a factor of the fine-structure constant squared ~$10^4$. Moreover, the QW structure is pumped electrically and a vertical doubly-resonant microcavity is designed to preferentially select the two-photon transition wavelengths modes: the signal $\omega_s$ and the idler $\omega_i$, by methods used for GaAs-based nonlinear optics [18]. Unlike the PDC-based sources [6], fundamental-wavelength photons (pump) are not required for this process – the energy is stored in the pumped material and one-photon emission may be suppressed. In an ideal TPE-based cavity-controlled source with a forbidden one-photon emission, majority of the injected carriers will recombine to emit a signal-idler photon pair, which has similar performance as an optically pumped PDC source with conversion efficiency near unity.

Hence theoretically, such a source should exhibit emission rates higher by many orders of magnitude compared to optically-pumped PDC-based sources at the same pump power levels. In practical electrically-driven semiconductor devices, however, such high pump powers are not feasible and at room-temperature the nonradiative recombination will be the dominant process and thus will determine the depletion rate of the carrier density in the QWs. Once the steady-state carrier density in the QWs at a given temperature is determined, the two-photon emission is determined as well, regardless of the residual one-photon emission rate. Therefore, for semiconductor devices operating at room-temperature the two-photon emission rate is determined solely by the carrier density. The photonic microcavity in such devices has a secondary role of reshaping the two-photon spectrum and emission direction, rather than overall rate enhancement and the suppression of the competing one-photon emission.

In QWs the light-hole (LH) heavy-hole (HH) degeneracy is removed, decoupling LH and HH levels for electron crystal momentum of k=0 and with proper strain the LH band can have the highest energy, while for non-vanishing k the LH and HH coupling is finite, dependent on k value and Luttinger parameters [11]. The vertical double-resonance cavity is designed to constrain the signal and the idler wavelengths to be emitted only by the two-photon transition between the zero-k conduction band (CB) subband-edge and the zero-k LH subband-edge which are the most populated states under regular injection conditions in such structures. Hence the angular momentum in the direction of the QW growth – z, in the valence band (VB) is [11] $j_z = \pm 1/2$ and CB $j_z = \pm 1/2$. Electron transitions between states of definite angular momentum will result in net angular momentum change of Δj$_z$= ±1 or Δj$_z$=0. The two-photon transitions must have therefore total angular momentum change of Δj$_z$=0. The cavity with higher reflectivity top mirror will launch the majority of photon pairs to be emitted collinearly downwards in the -z direction. Photon pairs emitted collinearly downwards will have therefore opposite polarizations and can be separated by a polarization beam splitter. The energy conservation for this two-photon process $\hbar(\omega_i + \omega_s) = E_{CB} - E_{VB}$ does not specify the energy of each individual photon and the emitted two-photon state is therefore energy-entangled.

$$|\Psi\rangle = \frac{1}{\sqrt{2}}\left(|\omega_i\rangle_R |\omega_s\rangle_L + |\omega_s\rangle_R |\omega_i\rangle_L\right) \quad (1)$$

where the entangled particles are identified by the polarization (Fig. 1). At room-temperature the electron angular momentum states will probably be mixed, however this does not impact the anti-correlated polarizations of the emitted photon pairs for polarization-based separation.

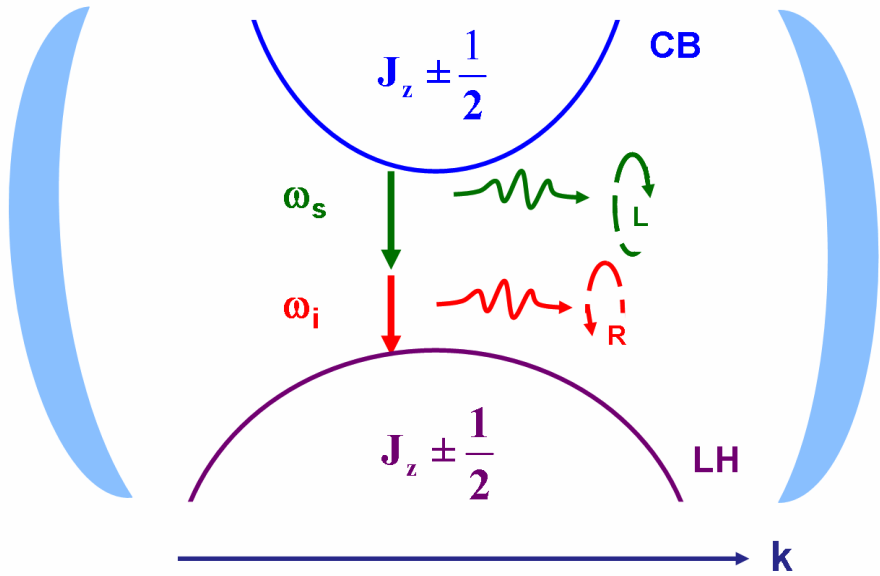

**Fig. 1 Cavity-controlled two-photon emission entanglement source**

Transitions between angular momentum superposition states in the VB and CB will result in total angular momentum change of $|\Delta j_z|>0$, corresponding to non-collinearly emitted photons with arbitrary polarizations. Only the photons emitted in the z direction can be tagged by polarization and separated by a polarization beam-splitter.

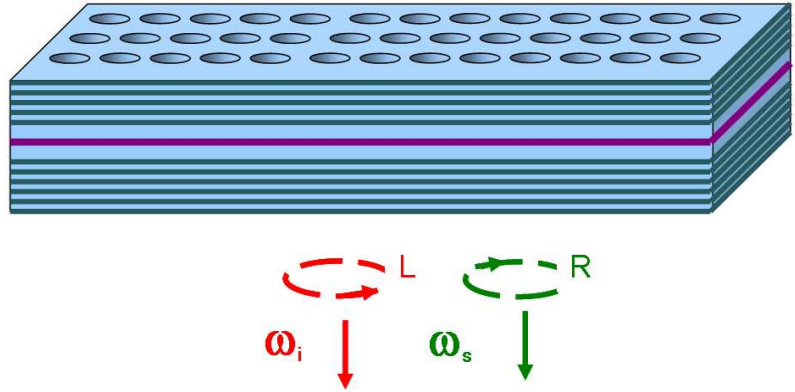

**Fig. 2 QWs in a vertical cavity combined with 2D photonic crystal, emitting polarization-tagged energy-entangled photons.**

In order to maximize the vertically propagating emission, a two-dimensional photonic crystal may be added (Fig. 2) [12], similar to the light extraction enhancement techniques used in light emitting diodes (LED). In addition, the Bragg type vertical cavity itself enhances (reshapes) preferentially the emission along the z direction by increasing the density of photonic states related to this emission. Similar photonic structures based on vertical cavity combined with shallow-corrugation two-dimensional gratings for enhanced-performance light-emitting diodes based on GaAs have been analyzed in detail [17] yielding extraction efficiencies as high as 40%. Application of such square-lattice grating to TPE emission wavelengths ~1.6μm results in a grating lattice period of ~490nm and filling factor of 0.5 wich are feasible using existing fabrication technology.

Cavity-controlled two-photon emission rate from a QW is calculated by a second order process in the time-dependent perturbation theory, similar to that of two-photon transition in a single atom [13]. The general expression for the rate is:

$$R = \frac{2\pi}{\hbar^2} N_e \int_{radiation\ states} F(\omega_1)F(\omega_2)d\omega_1 d\omega_2 \sum_{atomic\ states} |M|^2 \delta(\omega_0 - \omega_1 - \omega_2) \quad (2)$$

where $\hbar\omega_0$ is the bandgap energy, $N_e$ is the number of charge carriers and $F(\omega)$ is the density of radiation modes. M is given by:

$$M = \frac{N_c e^2}{2m_0 c^2}\left(\frac{2\pi\hbar c^2}{V}\right)\frac{1}{\sqrt{\omega_1 \omega_2}} M' \quad (3)$$

with electron charge e, free electron mass $m_0$ and the field quantization volume V, determined by the vertical cavity height and the unit cell of the horizontal 2D photonic crystal. $N_c$ is the number of photonic crystal unit cells for a specific device size and M' is the dimensionless matrix element for the second order process similar to that of the two-photon absorption [19]:

$$M' = \frac{2}{m_0}\sum_n \left(\frac{\langle f|\hat{p}\vec{\varepsilon_2}e^{-ik_2 x}|n\rangle\langle n|\hat{p}\vec{\varepsilon_1}e^{-ik_1 x}|i\rangle}{E_i - E_n - \hbar\omega_1} + \frac{\langle f|\hat{p}\vec{\varepsilon_2}e^{-ik_2 x}|n\rangle\langle n|\hat{p}\vec{\varepsilon_1}e^{-ik_1 x}|i\rangle}{E_i - E_n - \hbar\omega_2}\right) \quad (4)$$

where $\hat{p}$ is the momentum operator, $\vec{\varepsilon}$ is the photon polarization, i, n and f are the initial, intermediate and final electron states respectively.

Considering the initial state as the ground subband of the QW in the CB and the final state as the ground subband of the QW in the LH band, we base our calculations on a 14-band model [14] taking into consideration only the $p(\Gamma_7), p(\Gamma_8), p(\Gamma_8), \tilde{s}(\Gamma_6), \tilde{p}(\Gamma_7), \tilde{p}(\Gamma_8), \tilde{p}(\Gamma_8)$ with two spin states, with the higher conduction bands $\tilde{p}(\Gamma_7), \tilde{p}(\Gamma_8)$ taken as the intermediate states, while the other bands are neglected due to their energy remoteness.

Transitions between electrons states with high lattice momentum have larger matrix elements due to their k-dependence [11], however they are less populated by carriers. Nevertheless, the deterioration of the photon polarization anti-correlation due to these transitions can be minimized by the proposed vertical wavelength-selective cavity designed to match the k=0 transition.

The dipole matrix elements for $p(\Gamma_7), p(\Gamma_8), \tilde{s}(\Gamma_6)$ intermediate states are k-dependent, and being suppressed by the zero-k selecting cavity, will not contribute to the overall emission rate. Therefore, the two-photon transition probability can be calculated by using only the higher conduction bands $\tilde{p}(\Gamma_7), \tilde{p}(\Gamma_8)$ as intermediate states. Due to selection rules the envelope functions have equal quantum numbers, and the calculation of the matrix elements between the following states is performed using the dipole approximation:

$$\left|\frac{1}{2},\frac{1}{2}\right\rangle_{electrons} = |i\rangle = |S_\uparrow\rangle \cdot \phi_e(z) \cdot e^{j\vec{k}_e \cdot \vec{\rho}}$$

$$\left|\frac{3}{2},\frac{1}{2}\right\rangle_{light-hole} = |f\rangle = \left[\sqrt{\frac{2}{3}}|Z^v_\uparrow\rangle - \sqrt{\frac{1}{6}}|(X^v + iY^v)_\downarrow\rangle\right] \cdot \phi_{lh}(z) \cdot e^{j\vec{k}_{lh} \cdot \vec{\rho}}$$

$$|n_1\rangle = \sqrt{\frac{1}{2}}|X^c_\uparrow + iY^c_\uparrow\rangle \cdot \phi_{pe}(z) \cdot e^{j\vec{k}_{pe} \cdot \vec{\rho}}$$

$$|n_2\rangle = \left[\sqrt{\frac{1}{3}}|Z^c_\downarrow\rangle - \sqrt{\frac{1}{3}}|X^c_\uparrow - iY^c_\uparrow\rangle\right] \cdot \phi_{pe}(z) \cdot e^{j\vec{k}_{pe} \cdot \vec{\rho}}$$

(5)

where $\phi_i$ are the envelope functions, k is the inplane electron crystal momentum, ↑,↓ represent the spin state and X, Y and Z are the periodic parts of Bloch functions. The other pair of final and initial states (|f>=|3/2,-1/2> and |i>=|1/2,-1/2>) leads to the same rate and the same allowed photon polarizations. For the chosen intermediate states the infinite summation in Eq. 4 is replaced by only two non-vanishing different terms. Calculation of two-photon emission for x-polarized photons can be applied to the y-polarized emission as well, due to the z-axis rotation symmetry, and thus both right and left polarized photon transitions will have equal probabilities. The overlap of the different-band same-quantum-number envelope functions in infinite-barrier QW is unity, which is a good approximation for the QW structure used in our calculations. Using the non-centrosymmetric zincblende 14-band model matrix elements [14] yields:

$$M' = i\sqrt{\frac{3}{2}}\frac{P_1 Q}{m_e} \cdot \left( \frac{1}{E_c + \hbar\omega_s} + \frac{1}{E_c + \hbar\omega_p} - \frac{1}{E_c + \hbar\omega_s - \Delta_c} - \frac{1}{E_c + \hbar\omega_p - \Delta_c} \right) \quad (6)$$

where $P_1$ and $Q$ are the dipole moments for specific transitions, $E_{gap}$ is the QW energy gap, $E_c$ is the s-p bands energy difference and $\Delta_c$ is the higher conduction bands energy splitting.

For the inplane-propagating z-polarized two-photon emission the matrix element is 4 times larger than that of the vertical propagation; however it is suppressed by the proposed photonic crystal, whereas the two-photon transition with different i.e. inplane and vertical polarizations is totally forbidden. Assuming the cavity-controlled density of states $F(\omega)$ to consist of two well-separated Lorentzians:

$$F(\omega) = \frac{1}{2\pi}\left( \frac{\omega_s/(2Q_s)}{(\omega-\omega_s)^2 + (\omega_s/(2Q_s))^2} + \frac{\omega_i/(2Q_i)}{(\omega-\omega_i)^2 + (\omega_i/(2Q_i))^2} \right) \quad (7)$$

The calculated vertical emission rate is:

$$R = \frac{\pi^3 e^4 N_c}{m_0^2 V} n_e \frac{|M'|^2}{\omega_0 \omega_p \omega_s} \quad (8)$$

where $n_e = N_e/V$ is the charge carrier density.

Usually, in one-photon Purcell effect [15] the overall narrow-band one-photon emission, and hence the electron decay rate, are increased by modifying the cavity photon density of states to match the narrow-band free-space emission spectrum. In two-photon emission, however, the free-space emission spectrum is very wide-band, due to the different combinations of signal and idler wavelengths. For wide-band emitters placed in narrowband cavities the photon density of states has no effect on the overall emission intensity [16]. Therefore the cavity only determines the emitted spectrum shape and the angular distribution of the emission, but does not increase the overall electron decay rate. This is demonstrated by our calculated cavity-controlled two-photon emission rate, which does not depend on the cavity's quality factor (Eq. 8) and is similar to that of wide-band emission two-photon decay in free-space.

Assuming only the vertical emission is allowed by the photonic crystal, we calculate the pair generation rate for a structure of GaAs/AlGaAs QWs, injected carrier density of $n_e \sim 10^{19} cm^{-3}$ and a 1mm$^2$ device surface area. Vertical cavity size is chosen to be about half of the emitted-photon wavelengths, whereas further enlargement of the vertical cavity height would decrease the two-photon emission rate (Eq.8).

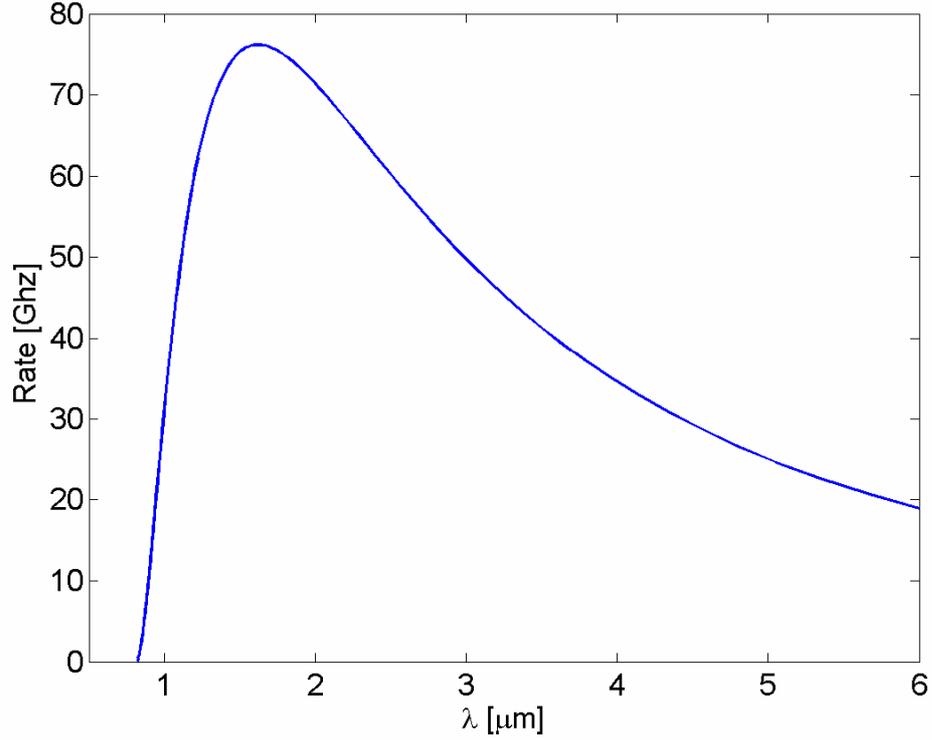

**Fig. 3 Entangled pair generation rate vs. one of the two-photon emission wavelengths.**

The maximal entangled pair generation rate for a half-wavelength vertical cavity size is $R \sim 7.5 \cdot 10^{10} sec^{-1}$ (Fig. 3). This rate corresponds to $\tau_{2ph} \sim 13ps$ average time interval between two-photon emission processes of and for a quality factor Q of 1000 the cavity photon lifetime is $\tau_{cav} \sim 2.4ps$. Higher emission rates for Q ~ 1000 might have caused accidental time-correlations between non-entangled photon pairs from different two-photon transitions thus reducing the entanglement generation rate. However for the calculated rate and Q, no more than one photon pair exists in the cavity at any given moment, preventing the deterioration of entanglement. The theoretically demonstrated narrow bandwidth room-temperature device rate is 3 orders of magnitude higher than that of the traditional broadband PDC based sources [8]. At lower temperatures with reduced nonradiative recombination rate, much higher emission rates are achievable.

In conclusion, we have demonstrated that a compact highly-efficient entangled-photons source can be realized via microcavity-controlled two-photon spontaneous emission from semiconductor QWs at room temperature. The microcavity is not expected to increase the total two-photon decay rate; nonetheless the desired wavelength emission is enhanced while the rest of the spectrum suppressed. The significantly more efficient second-order fundamental interaction, narrow bandwidth and collinear photon emission strongly enhance the efficiency leading to tens of GHz pair-generation rates, which are 3 orders of magnitude higher than those of the broadband PDC based sources. Furthermore, small dimensions, integrability, narrow bandwidth and room-temperature operation of this source may help introducing quantum information processing such as teleportation, quantum cryptography and quantum repeaters into the existing infrastructure of fiber-optical communications and integrated photonics

# References:

[1] C Bennett and S. J. Weisner, Phys. Rev. Lett. **69**, 2881 (1992).

[2] J. F. Clauser, M. A. Horne, A. Shimony, R. A. Holt, Phys. Rev. Lett. **23**, 880 (1969).

[3] A. Einstein, B. Podolsky, and N. Rosen, Phys. Rev. **47**, 777 (1935).

[4] A. R. Wilson, J. Lowe, and D. K. Butt, J. Phys. G. **2**, 613 (1976).

[5] A. Aspect, P. Gragnier, and G. Roger, Phys. Rev. Lett. **47**, 460 (1981).

[6] P. G. Kwiat et al. , Phys. Rev. Lett. **75**, 4337 (1995).

[7] R. Boyd, *Nonlinear Optics*, **2nd** edition (Academic Press, New York, 2003)

[8] M. Pelton, et al., Opt. Ex. **12**, 3573 (2004).

[9] X. Li, P. L. Voss, J. E. Sharping, P. Kumar, Phys. Rev. Lett. **94**, 053601 (2005).

[10] N. Akopian, N. H. Linder, E. Poem, Y. Berlatzky, J. Avron, D. Gershoni, B. D. Geradot, and P. M. Petroff, Phys. Rev. Lett. **96**, 130501 (2006).

[11] G. Bastard, *Wave Mechanics Applied to Semiconductor Heterostructures* (Les Editions de Physique, Les Ulis, (1988).

[12] K.Sakoda, *Optical Properties of Photonic Crystals* (Springer, Berlin, 2005 ).

[13] L. He and X. Feng, Phys. Rev. A **49**, 4009 (1994).

[14] M. E. Flatté, P. M. Young, L. -H. Peng, H. Ehrenreich, Phys. Rev. B **53**, 1963 (1996).

[15] C. Santori and et. al., Nature **419**, 594 (2002).

[16] R.Coccioli et al., Proc. Inst. Elect. Eng., **145**, 391 (1998).

[17] D. Delbeke, P. Bienstman, R. Bockstaele, and R. Baets, J. Opt. Soc. Am. B **19**, 871 (2002).

[18] G. Klemens, C. -H. Chen, and Y. Fainman, Opt. Express **13**, 9388-9397 (2005)

[19] C. C. Lee and H. Y. Fan, Phys. Rev. B **24**, 3502 (1974).